\documentclass{emulateapj}
\usepackage{apjfonts}

\newcommand{\msun}{{\rm \ M_\odot}}
\newcommand{\bd}[1]{{\bf{\textit{#1}}}}

\begin{document}
\title{The End of the MACHO Era:\\
Limits on Halo Dark Matter From Stellar Halo Wide Binaries}

\author{Jaiyul Yoo, Julio Chanam\'e, and Andrew Gould}

\affil{Department of Astronomy, The Ohio State University, 
140 West 18th Avenue, Columbus, OH 43210\\
jaiyul, jchaname, gould@astronomy.ohio-state.edu}

\slugcomment{accepted for publication in The Astrophysical Journal}

\begin{abstract}
We simulate the evolution of halo wide binaries in the presence of 
the MAssive Compact Halo Objects (MACHOs)
and compare our results to the sample of wide binaries of \citet{cg}.
The observed distribution is well fit by a single power law for angular
separations, $3.\!\!\arcsec5<\Delta\theta<900\arcsec$, whereas the simulated
distributions show a break in the power law whose location depends
on the MACHO mass and density. This allows
us to place upper limits on the MACHO density as a function of 
their assumed mass.
At the 95\% confidence level, we exclude MACHOs with masses $M>43\msun$
at the standard local halo density $\rho_H$. 
This all but removes the last permitted window for a full 
MACHO halo for masses $M>10^{-7.5}\msun$.
\end{abstract}

\keywords{dark matter --- Galaxy: halo --- methods: numerical --- 
stars: binaries --- stars: kinematics --- stars: statistics}

\section{Introduction}
After formation, wide binaries retain their original orbital parameters
except in so far as they are affected by gravitational encounters, or either
one or both binary members evolve off the main sequence.
These systems can therefore be used to probe inhomogeneities
of Galactic potential that may be due to black holes, low 
luminosity stars, molecular clouds, or other objects, because their low binding
energies are easily overcome by gravitational perturbations \citep{heg}.

\citet{bht} were the first to apply this principle, using wide binaries
in the Galactic disk to investigate disk dark matter 
in the Solar neighborhood.
\citet{martin} refined this approach by incorporating several effects that 
were previously ignored. In particular, they showed that gravitational 
encounters do not in general induce a sharp cutoff in the binary semi-major
axis distribution, but rather generate a break in its power law.
Although there have been extensive investigations of disk binary systems,
they have not yielded strong conclusions about the disk potential.
This is partly due to the relatively small size of the disk
binary samples available, partly to the fact that they do not have good
sensitivity for separations $\ga0.1$~pc, and partly because of the intrinsic
complexity of the disk potential. In principle, halo binaries could be
used to search for dark matter, but this has never been done, primarily
because there were no halo-binary samples adequate to the task.

However, halo dark matter has been investigated by several other techniques.
\citet{lo} suggested a mechanism for disk heating by supermassive black holes
and discussed that black holes with $M>10^6\msun$ could destroy the disk.
Null results of a search for  ``echoes'' of gamma-ray bursts 
induced by gravitational lensing constrained halo dark matter 
in the mass range  $M\sim10^{6.5}-10^{8.5}\msun$ \citep{ne1}. 
\citet{mo} argued that massive black holes $M>10^3\msun$
could disrupt low-mass globular clusters, which would imply an upper limit
$M<10^3\msun$. However, this argument is somewhat sensitive to assumptions
about the initial population of globular clusters.
Finally, microlensing experiments by the MACHO collaboration \citep{al4},
the EROS collaboration \citep{af}, and the two collaborations working together
\citep{al0} found that MAssive Compact
Halo Objects (MACHOs) with $10^{-7.5}\msun<M<30\msun$ cannot 
account for the mass of the dark halo.
\citet{der} argued that baryonic dark matter with $M<10^{-7}\msun$ 
would have evaporated away in a Galactic time scale.

The publication of the $\mu>0.\!\!\arcsec18{\rm \ yr^{-1}}$ proper motion
limited New Luyten Two Tenths (NLTT) catalog \citep{nltt1,nltt2}
has vastly increased the pool of available data on binary systems,
but the short color baseline of the photographic photometry
in NLTT rendered halo/disk discrimination extremely difficult.
However, \citet{sa1} and \citet{sa2} revised NLTT (rNLTT) with improved
astrometry and photometry for the 44\% of the sky covered by the intersection
of the Second Incremental Release of the Two Micron All Sky Survey
and the first Palomar Observatory Sky Survey. 
Chanam\'e \& Gould (2003, hereafter CG)
then constructed a homogeneous catalog of binaries
from rNLTT and classified each entry as either disk or halo.

In this paper, we use Monte Carlo simulations to evaluate the effect
of MACHOs on the halo binary distribution and compare these predictions
with the observed halo binary sample of CG by means of a
likelihood analysis.
We briefly describe our CG halo binary sample in \S~\ref{wb} 
and make simple analytic estimates of the effects of perturbers in 
\S~\ref{te}. Detailed justifications of our assumptions and our 
Monte Carlo algorithm are presented in \S~\ref{ns}. 
We present the results of our numerical
simulations in \S~\ref{re} and the likelihoods of halo dark-matter models
in \S~\ref{sw}. Finally, we summarize our results in \S~\ref{co}.

\section{The CG Halo Wide Binaries}
\label{wb}
\begin{figure}[t]
\centerline{\epsfxsize=3.5truein\epsffile{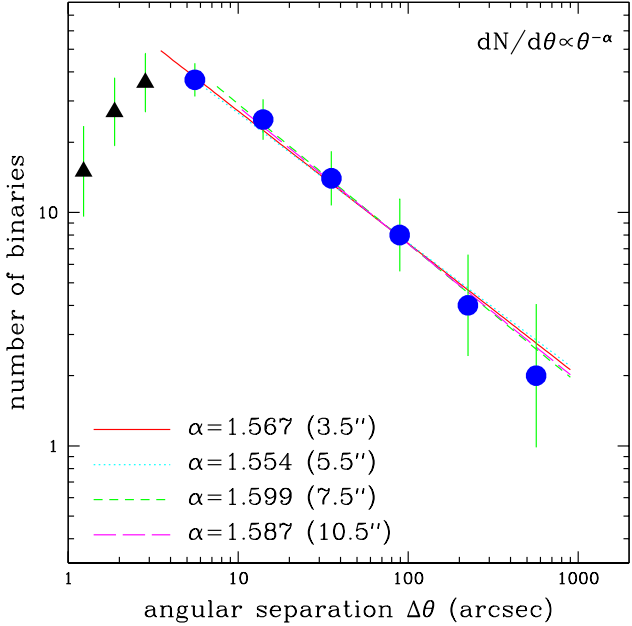}}
\caption{Halo binary distribution function from the catalog of \citet{cg}.
Samples of wide binary systems are complete up to 900\arcsec~while the lower
limit in angular separation is not precisely established.
The circles represent actual counts in six equal logarithmic bins over
$3.\!\!\arcsec5\leq\Delta\theta<900\arcsec$. The triangles show three
$1\arcsec$ bins for $\Delta\theta<3.\!\!\arcsec5$ and are rescaled to
account for the smaller bin sizes relative to the circles.
Various lines represent fits for subsamples with different lower limits 
in angular separation. Error bars represent one standard deviation 
$\sigma=n^{-1/2}/\ln10$.}
\label{data}
\end{figure}
Figure~\ref{data} shows the halo binaries from the CG catalog upon which the 
current work is based. This binary distribution function 
is well approximated by a single power law from 
$\Delta\theta=3.\!\!\arcsec5$ to the
catalog limit at $\Delta\theta=900\arcsec$. However, since the lower threshold
of completeness is not precisely established, we consider several different
lower limits. 
There are respectively (90, 68, 59, 47) binaries in the samples with lower
limits ($3.\!\!\arcsec5$, $5.\!\!\arcsec5$, $7.\!\!\arcsec5$, 
$10.\!\!\arcsec5$). For $\Delta\theta\lesssim3.\!\!\arcsec5$, the sample 
is not complete, and the deviation from a power-law distribution is clearly
shown as triangles in Figure~\ref{data}. 
However, since the fit is insensitive to the 
choice of lower limit ($\Delta\theta\geq3.\!\!\arcsec5$), we adopt
a lower threshold of $\Delta\theta=3.\!\!\arcsec5$. 
Since severe incompleteness clearly sets in at $\Delta\theta<3.\!\!\arcsec5$,
one might be concerned about more modest incompleteness just above our
adopted boundary $\Delta\theta=3.\!\!\arcsec5$. However, as we discuss
in \S~\ref{sw}, incompleteness near $\Delta\theta\sim3.\!\!\arcsec5$
would act to lessen the likelihood difference between a halo dark-matter 
model and the observations. Therefore, our procedure is conservative.

Assuming that the initial binary distribution is characterized by a power
law, the slope of which might be different from the final observed one,
the observed binary distribution can yield significant constraints on 
halo dark matter. Since binary systems with large semi-major axes,
$a\ga0.1$~pc, are more liable to
be disturbed by perturbers than those with small $a$, the deviation 
(or lack thereof) from a power-law distribution at wide angular separations
provides a test of halo dark-matter models.

\section{Theoretical Expectations}
\label{te}
We begin by considering two extreme regimes; the tidal regime 
($b_{\rm min}\gg a$) 
and the Coulomb regime ($b_{\rm min}\ll a$), where $a$ is the binary 
semi-major axis and,
\begin{eqnarray}
b_{\rm min}&=&\left({M\over\pi\rho v T}\right)^{1/2} \nonumber \\
&=&700{\rm \ AU}
\left({M\over{\rm M}_\odot}\right)^{1/2}\left({\rho\over\rho_H}
\right)^{-1/2}\left({v\over{300\rm \ km\ s^{-1}}}\right)^{-1/2}
\label{bmin}
\end{eqnarray}
is the typical minimum impact parameter for perturbers 
of mass $M$, density $\rho$,
and velocity $v$ over the duration $T$ that the binary is subjected to
perturbations. We evaluate 
equation~(\ref{bmin}) using the standard local halo density
$\rho_H=0.009\msun{\rm \ pc^{-3}}$ \citep{lh}. 
We adopt $T=10$~Gyr. Note that while halo stars (and so halo binaries) are 
several Gyr older than this, $T$ represents the time since the binaries
were assembled into the Galactic halo and not the time since their formation.

In the tidal regime, the perturbations are dominated by the single closest
encounter, which yields a change of relative velocity,
\begin{equation}
\Delta\bd{v}\simeq{2GM\over b^2v}a.
\end{equation}
By evaluating $\Delta\bd{v}$ at $b=b_{\rm min}$ and
equating the result with the internal velocity of the binary system, 
$v^2=Gm/a$, 
we obtain an estimate of transition separation,
\begin{equation}
a_t=\left({m\over4\pi^2G\rho^2T^2}\right)^{1/3}
\simeq18,000{\rm \ AU}\left({\rho\over\rho_H}\right)^{-2/3},
\label{tran}
\end{equation}
where we adopt $m=1\msun$ as the typical total mass of the binary system.
The corresponding transition orbital period is 
$P_t\simeq2{\rm \ Myr}~(\rho/\rho_H)^{-1}$.

Binary systems whose semi-major axes are less than $a_t$ or equivalently
whose
orbital periods are shorter than $P_t$ are insensitive to perturbations.
Note that the transition separation in the tidal regime is 
independent of the mass and velocity of perturbers.
This regime approximately applies when $b_{\rm min}\gtrsim a_t$, i.e.,
\begin{equation}
M\gtrsim\left({m^2v^3\over16\pi G^2\rho T}\right)^{1/3}
\simeq700\msun.
\label{mcrit}
\end{equation}

In contrast to the tidal regime, which is dominated by the 
single closest encounter, the perturbations in the Coulomb regime 
(named for the Coulomb logarithm, $\ln\Lambda\equiv\ln(a/b_{\rm min})$),
are described by continuous weak gravitational encounters,
\begin{equation}
(\Delta v)^2=2\int_{b_{\rm min}}^a\!\!\!\!2\pi b~{\rm d}b 
\left({2GM\over bv}\right)^2{\rho\over M}vT
={16\pi G^2\rho MT\over v}\ln\Lambda,
\label{eq:cou}
\end{equation}
where the factor 2 in front accounts for the independent perturbations of
each component of the binary. The transition separation is then,
\begin{equation}
a_t={mv\over16\pi G\rho MT\ln\Lambda}=
{3000{\rm \ AU}\over\ln\Lambda}\left({M\over10^3\msun}\right)^{-1}
\left({\rho\over\rho_H}\right)^{-1},
\label{cou}
\end{equation}
and so is roughly inversely proportional to the perturber mass.

\section{Numerical Simulation}
\label{ns}
In this section, we summarize our Monte Carlo algorithm, which evaluates
the effect of all perturbations using a simple impulse approximation
and ignores such effects as large-scale tides and molecular clouds.
We include the contribution from the ionized binaries,
although as we discuss in \S~\ref{sec:ion} and show more fully in the Appendix,
disrupted binaries diffuse to separations well beyond the observational range
of interests in a time very short compared to $T$.

Since previous work, notably that of \cite{martin}, has focused considerable
effort on incorporating the above-mentioned effects, 
we first justify our decision
to ignore them.  It is primarily the fact that we are considering halo
binaries whereas previous workers were investigating disk binaries
that accounts for the difference in importance of these effects.

\subsection{Impulse Approximation}
In the Coulomb regime, the biggest relevant impact parameter is of order $a$. 
Since the perturbers are moving much faster than the binary components,
the impulse approximation well describes encounters.

In the tidal regime,
the perturbations are dominated by the closest single encounter,
whose crossing time is,
\begin{eqnarray}
T_c&=&{2b_{\rm min}\over v} \nonumber \\
&=&20{\rm \ yr}\left({M\over\msun}\right)^{1/2}
\left({\rho\over\rho_H}\right)^{-1/2}\left({v\over300{\rm \ km\ s^{-1}}}
\right)^{-3/2}.
\end{eqnarray}
Since, for masses $M\lesssim10^8~M_\odot$,
the crossing time is significantly less than the transition
orbital period $P_t\simeq2{\rm \ Myr}$,
the impulse approximation is always justified.

\subsection{Disk and Galactic Tides}
When halo binaries pass through the plane of Galactic disk, differential
gravitational attractions give rise to tidal effects, also known as disk
shocking \citep{gd}. 
Since the mean velocity at which halo binaries cross the disk is
$v_z\sim100{\rm \ km\ s^{-1}}$ \citep{pg1,pg2}, the change in the
z-component of velocity of halo binaries is
$\Delta v_z\simeq4\pi G\Sigma a/v_z$
where $\Sigma=40\msun{\rm \ pc^{-2}}$ is the surface density of the Galactic
disk in the Solar neighborhood (\citealt{zh}, and references therein).
Equating this to the internal velocity of the binary system, $v^2=Gm/a$, 
yields,
\begin{equation}
a_{\rm crit}=\left({mv_z^2\over16\pi^2 G\Sigma^2}\right)^{1/3}
\simeq2{\rm \ pc},~~~~~({\rm Disk~Tides}).
\end{equation}

For the most favorable (i.e., prograde) orbits, 
the binary system can be disrupted
by Galactic tidal fields if the internal orbital period is longer than
a Galactic year, $P_G\simeq230{\rm \ Myr}$.
The critical semi-major axis is therefore,
\begin{equation}
a_{\rm crit}={\rm AU}\left({m\over\msun}{P_G^2\over{\rm yr^2}}\right)^{1/3}
\simeq2{\rm \ pc},~~~({\rm Galactic~Tides}).
\end{equation}

Since these critical values are not much larger than the largest projected
separation in our sample, one might be concerned that
the effects of disk and Galactic tides are non-negligible.
However, the observational data show that the binary distribution is well
described by a single power law within the limits of
$3.\!\!\arcsec5<\Delta\theta<900\arcsec$,
and we are being conservative to ignore
both tides because incorporating their effects 
would only serve to place more stringent constraints
on the perturbers. On the other hand, if the binary distribution had shown
a power-law break, which would have been indicative of dark matter, 
it would have been necessary to investigate how much of this 
signature was actually due to tides.

\subsection{Giant Molecular Clouds}
With typical masses $10^5-10^6\msun$ and surface density
$\Sigma\sim5\msun{\rm \ pc^{-2}}$, corresponding to a mean density
$\rho\sim6\times10^{-4}\msun{\rm \ pc^{-3}}$ over typical halo
orbits, giant molecular clouds (GMCs) are clearly in the tidal regime.
Because of their low density $\rho\sim 0.07\rho_H$, they would
add very modestly to the effect of halo tidal perturbers.  
The effect of GMCs is further diminished by the fact that $b_{\rm min}$
(eq.~[\ref{bmin}]) is substantially smaller than a typical cloud 
(and this remains so even if one considers the individual ``clumps''
inside a GMC). Hence, we ignore GMCs.  
As with tides, the effect of doing so is conservative:
taking account of GMCs would only strengthen any limits we obtain.

\subsection{Ionized Binaries}
\label{sec:ion}
Even though binary systems are disrupted by perturbers, the stars of former
binary members escape the system with extremely low escape energies due to
the interplay  of diffusive processes and the tidal boundary rather than
disappearing immediately after disruption.
As shown by the spectacular images of the globular cluster Palomar~5 suffering
on-going tidal disruptions \citep{pal5}, the mass in a tidal tail
can be larger than the mass in the parent body. By analogy, ionized binaries
could in principle stay well inside the observational range of interest.

However, globular cluster escapees are no longer perturbed and so proceed
on deterministic orbits that separate from the cluster at a roughly uniform
rate, while the binary members drift away in the Galactic potential
and continue to be heated by the omnipresent perturbers after they are 
well enough separated to be free from each others' gravitational influence.

In the Appendix, we quantify this argument and analytically show that
ionized binaries do not contribute significantly to the observed distribution.
We also give a prescription for including them in our simulations, and
indeed we find that they contribute negligibly.

\subsection{Non-Circular Orbits}
In our simulations, we assume that the binaries are subjected to perturbations
by ambient objects whose density is constant in time.
Strictly speaking this would apply only to binaries in circular Galactic
orbits. More typical binaries would tend to spend the majority of their
time farther from the Galactic center than the Sun, where the perturber
density is lower. However, we find by numerically integrating a wide
range of orbits, that the average density of perturbers encountered
by binaries in non-circular orbit is almost always higher than for those
in circular orbits. Hence, our approximation of constant perturber density
is conservative.
\begin{figure}[t]
\centerline{\epsfxsize=3.5truein\epsffile{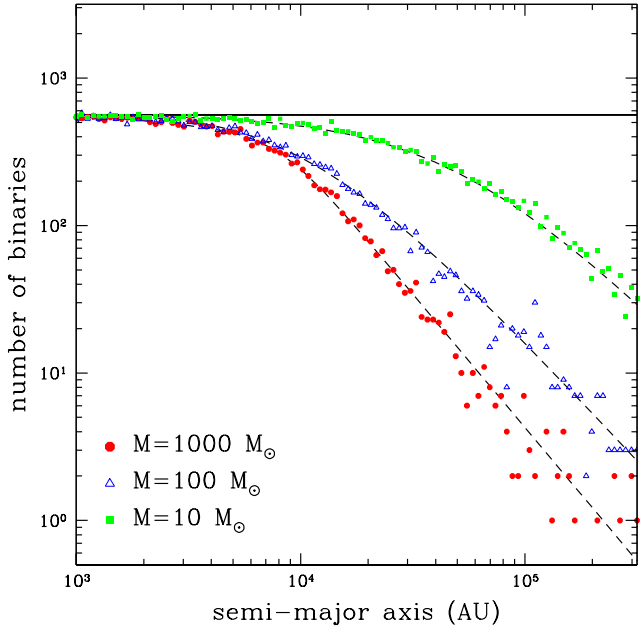}}
\caption{Binary distributions as a function of semi-major axis. 
100,000 binaries are generated following an arbitrarily chosen
flat ($\alpha$=1) distribution represented as a thick solid line. 
The halo density is set to be $\rho_H$. The squares, triangles 
and circles represent binary distributions for three different masses of 
perturber, after $T=10$ Gyrs evolution. The fitting curves for each model are 
shown as dashed lines.}
\label{flat}
\end{figure}

\subsection{Monte Carlo Algorithm}
\label{monte}
To test the consistency of the observed binary distribution with the presence
of massive perturbers, we must be able to simulate the effect of these
perturbers on arbitrary power-law initial distributions. Our basic method
for doing this is Monte Carlo simulations in which the initial binary 
semi-major axis is drawn uniformly from power-law distribution over the
interval, $1<\log(a/{\rm AU})<5.5$. 
The eccentricity is drawn randomly 
from a distribution uniform in $e^2$ \citep{martin},
and the phase and orientation of the
orbit are assigned randomly.
The perturbers are assumed to have an isotropic Maxwell\-ian velocity
distribution relative to the binary center of mass, with a one-dimensional
dispersion $\sigma=200{\rm \ km\ s^{-1}}$.  
This reproduces the true rms velocity, which
is a combination of an isothermal-sphere perturber distribution of circular
speed $v_c=\sqrt{2}\sigma=220{\rm \ km\ s^{-1}}$, and the measured dispersions
of halo stars\footnote{\citet{cb} obtained dispersions of halo stars
$(\sigma_\pi,\sigma_\theta,\sigma_z)=$ $(141,106,95)~{\rm km\ s^{-1}}$.
The corresponding one-dimensional dispersion for perturbers
is $\sigma=194~{\rm km\ s^{-1}}$, which is therefore
insensitive to the precise choice of $(\sigma_\pi,\sigma_\theta,\sigma_z)$.} 
$(\sigma_\pi,\sigma_\theta,\sigma_z)=$ 
$(170,97,93)~{\rm km\ s^{-1}}$ \citep{pg1,pg2}.
In principle, the velocity distribution should be treated as anisotropic.
However, this is a higher-order
effect, which we ignore in the interest of simplicity.

We consider all impact parameters with 
$b\lesssim b_{\rm max}={\rm max}\{10b_{\rm min},2a\}$. 
In particular, we evaluate 
the $\sim100$ closest
impacts in the tidal regime, even though as we discuss in \S~\ref{te},
the single closest encounter dominates.
The mass of each binary component is set to be $0.5\msun$.
We evolve 100,000 binary systems in each simulation.
To illustrate the dependence on the mass of perturbers, we show their
effects on an artificial initial binary distribution that is independent 
of semi-major axis (see Fig.~\ref{flat}).
As expected, for the binary systems that are initially tightly bound,
the final distributions are almost the same as the 
initial ones regardless of the mass of the perturbers.
However, at wide separations, the distributions are driven to a new
power law, which becomes steeper with increasing mass.

This approach works well for any individual initial power-law distribution, but
is too time-consuming to process the very large number of distributions
required for the comparison of data and models. In \S~\ref{sm}, we introduce
a scattering-matrix formalism that substantially improves the efficiency
of mass-production Monte Carlo simulations.

\subsection{Fitting Formula}
Motivated by the fact that the final distribution is 
well-approximated by power laws at each extreme,
we use a two-line (five-parameter)
fitting formula given by,
\begin{equation}
H(x)=\left[f(x)^{-n}+g(x)^{-n}\right]^{-1/n},
\label{fit}
\end{equation}
where $f(x)$ and $g(x)$ are (two-parameter) straight lines in their argument,
$x=\log(a/{\rm AU})$, each
corresponding to the respective asymptotic behaviors of $H(x)$.
The fifth parameter $n$ permits a smooth transition between
$f(x)$ and $g(x)$ in the intermediate region.
We calculate the five parameters of equation~(\ref{fit}) by minimizing
$\chi^2$ for a given data set, and the fitting curves for four different 
masses of perturber are represented as dashed lines in Figure~\ref{flat}.
\begin{figure}[b]
\centerline{\epsfxsize=3.5truein\epsffile{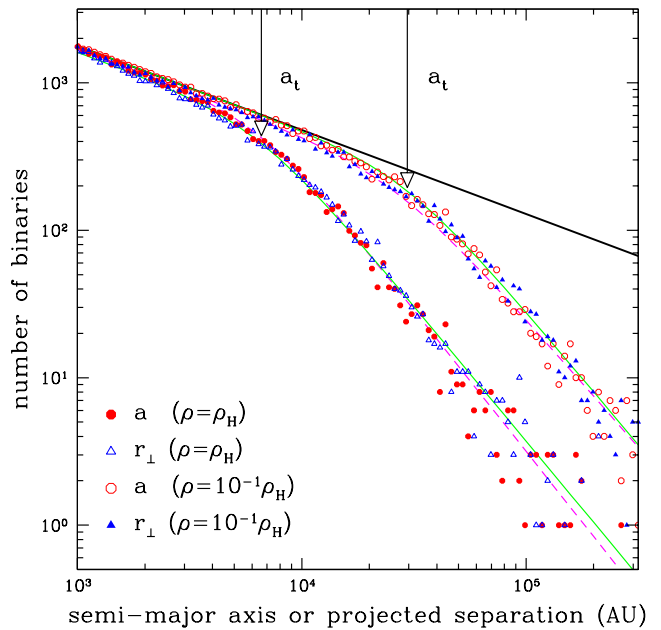}}
\caption{Evolution of binary distributions with the initial power law,
$\alpha=1.567$ obtained in \S~\ref{wb}, represented as a thick solid line.
50,000 binaries are evolved in the presence of perturbers of mass 1000$\msun$ 
The circles show the binary distributions as functions of semi-major 
axis while the triangles show the distributions of projected physical 
separations of the binary components onto the sky plane.
The solid lines represent fitting curves for semi-major axis, and
the dashed lines for projected physical separation. Two vertical arrows 
indicate transitions, $a_t$, from the unperturbed to the perturbed regimes.}
\label{pow}
\end{figure}

\section{Results}
\label{re}
Here, we present our main calculations on the evolution of binary distributions
with various initial slopes
under the influence of various perturber masses and halo densities,
and we evaluate the transition separations.
Although the semi-major axis of a binary system is a direct indicator of the 
binding energy of the system and is the theoretically most tractable quantity,
it is not observable. It is the angular
separations on the sky that we can directly measure from observations.
To compare our results with the data, we calculate physical 
separations projected on the sky plane and convolve these with an adopted
distance distribution to predict the binary distributions
as a function of angular separation.

In principle, one could compare models directly to the observed
projected separations since CG give individual distance
estimates to each binary.  However, while the observational selection
function is quite simple for angular separations (essentially just
a pair of $\Theta$-functions), it is rather complex and would be
extremely difficult to model for projected physical separations.
Hence, we compare our models to the most directly observed quantity:
angular separations.

\subsection{Semi-Major Axis, $a$}
We begin the simulation with a flat distribution,
$dN/d\log a\propto const$ for calculation of scattering matrices.
We then investigate the dependence of the resulting
final binary distribution on perturber mass and halo density. 
Figure~\ref{pow} shows a sample of our results with the initial power-law
distribution, $dN/d\log a\propto a^{-0.567}$, in agreement with the 
observations as summarized in \S~\ref{wb}.
Two models with the same perturber mass but widely different halo densities,
$\rho_H$ and $10^{-1}\rho_H$ are shown. The distributions of semi-major axis 
are represented by circles.

\subsection{Projected Physical Separation, $r_\perp$}
At $T=10$~Gyr, we calculate the projected physical separation,
$r_\perp$ of each binary taking account of the randomly chosen 
orientation to the line of sight and the orbital phase.
The distributions of $r_\perp$ are shown by triangles in Figure~\ref{pow}.

The two sets of distributions are very similar to each other
as demonstrated by the solid and
the dashed lines, which are fitted to the points.
This can be understood as follows. The time-averaged physical separation is
\begin{equation}
\left<r(e)\right>={1\over2\pi}\int_0^{2\pi}a(1-e~{\rm cos}~\psi)^2{\rm d}\psi
=a\left(1+{e^2\over2}\right).
\end{equation}
Averaging over the uniform distribution in $e^2$ yields 
$\left<r\right>=5a/4$, and projecting onto the sky plane gives,
\begin{equation}
\left<r_\perp\right>={\pi\over4}\left<r\right>={5\pi\over16}a\simeq0.98a.
\label{phy}
\end{equation}
That is, the triangles are, on average, slightly shifted to left of the circles
in Figure~\ref{pow}. 

There is, however, a deviation from equation~(\ref{phy}) at large projected 
physical separation. We find that a binary system with a highly eccentric 
orbit is more likely to be disturbed than one with a circular orbit.
Since the highly eccentric systems are selectively destroyed for
loosely bound binary systems, the distribution in eccentricity is 
no longer isotropic, and the actual projected physical separations differ from
equation~(\ref{phy}).
We therefore use numerical calculations of the projected physical
separations rather than the analytic estimate of equation~(\ref{phy}).

\subsection{Transition Separation, $a_t$}
Using the five-parameter fitting formula from equation~(\ref{fit}),
we calculate the asymptotic slopes for distributions in semi-major axis.
These are represented as solid curves in Figure~\ref{pow}.
The intersection of these asymptotic slopes characterizes the transition $a_t$
from the unperturbed to the perturbed regime. 
The two arrows in Figure~\ref{pow} 
represent the intersections for two different halo densities. 
As predicted by equation~(\ref{tran}), the transition for a denser
halo model occurs at smaller separation, $a_t\propto\rho^{-2/3}$.

\begin{figure}[t]
\centerline{\epsfxsize=3.5truein\epsffile{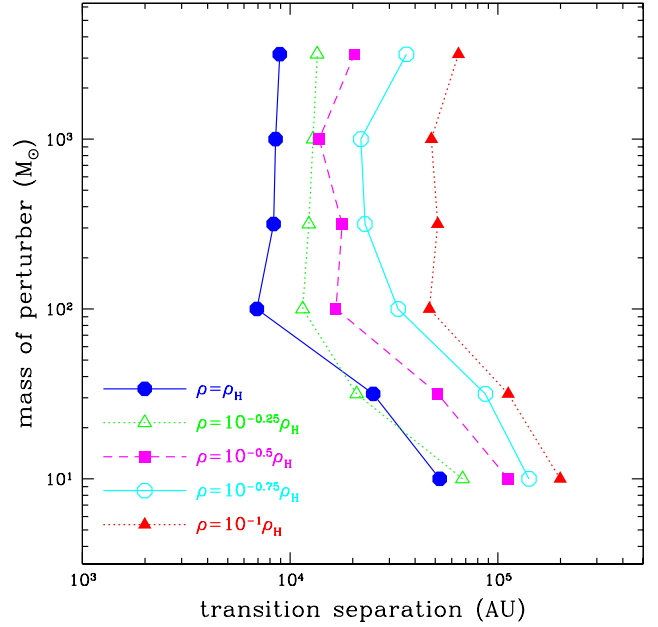}}
\caption{Transition separations $a_t$ for different halo densities as 
a function of perturber mass. Transition separations are obtained for 
each model by calculating the intersection of $f(x)$ and $g(x)$ in 
eq.~(\ref{fit}).}
\label{int}
\end{figure}

The transition separations for various perturber mass and 
halo densities are shown in Figure~\ref{int}, which provides a 
qualitative understanding of the underlying physics of binary disruption
and is in a good accord with the prediction of 
equation~(\ref{tran}) and (\ref{cou}).
In particular, in the tidal regime the transition separation 
is independent of perturber mass, and is a function of halo density 
$a_t\propto\rho^{-2/3}$,
although the actual values of $a_t$ are a factor of $2$ smaller than
those predicted by equation~(\ref{tran}).
At smaller mass, the transition separation grows roughly as
$a_t\propto M^{-1}$, as predicted by equation~(\ref{cou}) for the Coulomb
regime.

\subsection{Angular Separation, $\Delta\theta$}
\label{as}
To predict the observed angular distribution function, we
must convolve the model's projected-separation distribution function
with an assumed distribution of (inverse) distances to the binaries in the
sample.  

We derive this distribution from the observed distances of the 90
binaries in the CG sample.  To understand our procedure for doing
so, consider first the idealized case of binaries drawn from a
distance-limited sample of uniform density.  Of course, in this case,
the distance distribution would be 
$d N/d \log R \propto R^3 \Theta(R_{\rm max}-R)$,
where $\Theta$ is a step function, and $R_{\rm max}$ is the distance limit.   
However, the binaries selected
at fixed $\Delta\theta$ would not be distributed as $R^3$.  This  is
because the binaries at distances $R_1$ and $R_2$ would have projected
separations $r_{\perp,1}= R_1\Delta\theta$ and
$r_{\perp,2}= R_2\Delta\theta$, and these differ in frequency (for fixed log
intervals) in the ratio $(r_{\perp,1}/r_{\perp,2})^{1-\alpha}$, where 
$\alpha=1.567$ is the
binary-separation power-law slope.  Hence, the sample distribution
would be $d N_{\rm sample}/d\log R \propto R^{4-\alpha} \Theta(R_{\rm max}-R)$.
Therefore, to obtain an estimate of the underlying distribution of distances
that yields the observed sample distribution, we simply adopt the
observed distribution, but weight each binary by $R^{\alpha-1}$.
We estimate $R$ for each binary by applying the fourth-order color-magnitude
relation of CG to the brighter component, except when that
component has no $J$-band data, in which case we use the fainter component.

\begin{figure}[t]
\centerline{\epsfxsize=3.5truein\epsffile{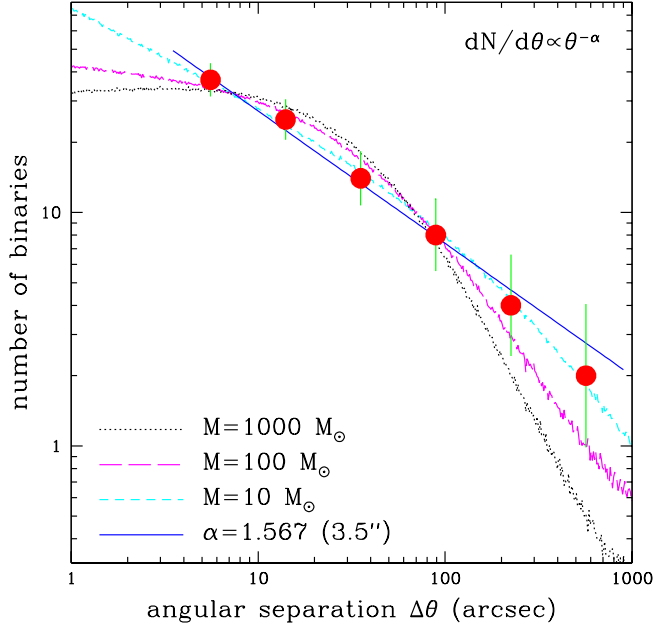}}
\caption{The best-fit final binary distributions for various perturber masses,
assuming that the initial distribution is a power-law.
The halo density is set to be $\rho_{\rm H}$. The observed halo binary 
distribution (Fig.~\ref{data}) is shown for comparison.
A model with 1000~$M_\odot$ perturber deviates significantly from the 
observations while a model with 10~$M_\odot$ is quite consistent with them.}
\label{add}
\end{figure}

\subsection{Scattering Matrix, $\bf{S}$}
\label{sm}
To simultaneously investigate large subsets of initial power-law 
distributions for each
model specified by $M$ and $\rho$, we use a Monte Carlo simulation 
(see \S~\ref{monte}) with an initially flat ($dN/d\log a\propto const$)
distribution to construct a scattering matrix, ${\bf{S}}_{ij}(M,\rho)$, 
the relative probability that a binary with initial log semi-major axis,
$\log a_j$ will finally be observed in the angular separation bin,
$\log \Delta\theta_i$. Explicitly,
\begin{equation}
{\bf{S}}_{ij}(M,\rho)=\sum_k\int{\rm d}\log R{dN\over d\log R}
\delta\left[\log R\Delta\theta_i-\log r_{\perp,j}(k;M,\rho)\right],
\label{seq}
\end{equation}
where $r_{\perp,j}(k;M,\rho)$ is the final projected separation of a binary
with initial semi-major axis $a_j$ in $k^{th}$ realization of a Monte Carlo
simulation of perturbers with mass $M$ and density $\rho$.
In practice, since the radial profile $dN/d\log R$ is estimated directly from
the binary data as discussed in \S~\ref{as}, we evaluate  
equation~(\ref{seq}) by discretely summing over the $l=1\ldots90$ binaries,
\begin{eqnarray}
{\bf{S}}_{ij}&(M,\rho)&=\sum_k\sum_{l=1}^{90}R_l^{\alpha-1} \\
&&\times\Theta
\left({\delta\log\Delta\theta\over2}-|\log R_l\Delta\theta_i-\log r_{\perp,j}
(k;M,\rho)|\right),  \nonumber
\end{eqnarray}
where $\delta\log\Delta\theta$ is the logarithmic width of the $\Delta\theta$
bins. 

Since the simulations are calculated
from a flat distribution, i.e., equal number of binaries
per logarithmic semi-major axis bin, the un-normalized probability of finding
binaries at $\Delta\theta$ for a given dark-matter model is then given by,
\begin{equation}
P(\Delta\theta_i;M,\rho,\alpha)={\bf{S}}_{ij}(M,\rho)w(a_j;\alpha),
\end{equation}
where $w(a_j;\alpha)=a_j^{1-\alpha}$ is the power-law of the
initial binary distribution, $dN/da\propto a^{-\alpha}$.

\section{Dark Matter Limits}
\label{sw}
For each model, we fix the normalization so that
the number expected in the interval 
$3.\!\!\arcsec5<\Delta\theta<900\arcsec$ is equal to
90, i.e., the total number observed over this interval.  We call the
normalized resulting function 
$P_{\rm N}(\Delta\theta;M,\rho,\alpha)$, evaluate the likelihood
of a given model as a sum over the 90 observed binaries, 
\begin{equation}
\ln L(M,\rho,\alpha)=\sum_{i=1}^{90}\ln P_{\rm N}(\Delta\theta_i;M,\rho,
\alpha),
\label{eqn:likelihood}
\end{equation}
and find the maximum likelihood over various initial slopes $\alpha$,
\begin{equation}
\ln L(M,\rho)\equiv\max_\alpha\left\{\ln L(M,\rho,\alpha)\right\}.
\end{equation}
\begin{figure}[b]
\centerline{\epsfxsize=3.5truein\epsffile{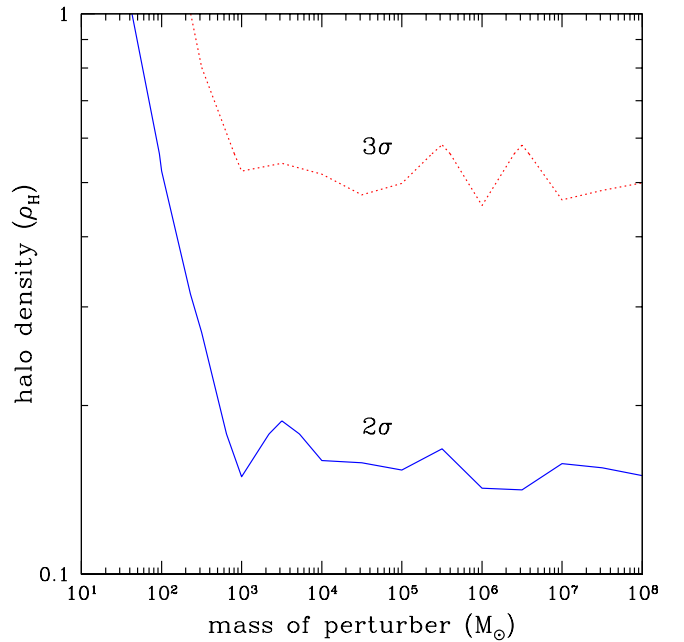}}
\caption{Exclusion contour plot for halo dark-matter models. 
Various confidence levels are shown. The oscillations at high masses are
due to numerical noise.}
\label{con}
\end{figure}
\begin{figure}[t]
\centerline{\epsfxsize=3.5truein\epsffile{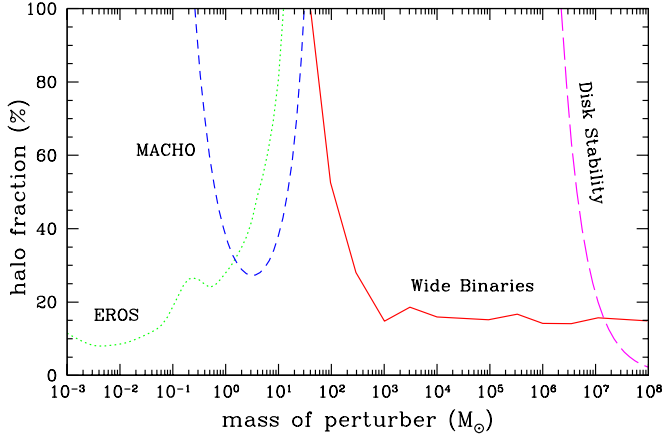}}
\caption{Exclusion contour plot at 95\% confidence level.
The dashed, the dotted and the long dashed lines represent the 
microlensing-based limits from EROS \citep{af} and MACHO \citep{al4}, and the 
limit based on disk stability \citep{lo}, respectively. Our limits from the
$3.\!\!\arcsec5<\Delta\theta<900\arcsec$ sample are 
represented as a solid line.}
\label{comb}
\end{figure}

Figure~\ref{add} shows the normalized resulting functions for various 
perturber masses with $\rho=\rho_{\rm H}$. For 1000 $M_\odot$ perturbers,
the likelihood is maximized by an initial distribution that is nearly flat.
However, no initial power-law is consistent with the observed distribution.
For low mass perturbers, by contrast, the likelihood is maximized by an initial
slope that is only slightly shallower than the observed one, and the resulting
distribution is consistent with observations.

We compare the resulting likelihoods to $L(0,0)$, i.e., the model with
no dark-matter perturbers, which is characterized by the best fit slope
$\alpha=1.567$, and we define confidence levels by
\begin{equation}
\sigma(M,\rho) =\sqrt{2\left[\ln L(0,0)-\ln L(M,\rho)\right]}.
\label{eqn:sigmas}
\end{equation}

As we discussed in \S~\ref{wb}, the fact that the binary sample is clearly
incomplete for $\Delta\theta<3.\!\!\arcsec5$ implies that it may also
be slightly incomplete for binaries just above this threshold. However,
one can see from the form of the normalized resulting functions in
Figure~\ref{add} that such incompleteness would actually favor models
with massive perturbers over the no-perturber model. We verify numerically
that this indeed is the sign of the bias. Hence our choice of the 
$\Delta\theta\geq3.\!\!\arcsec5$ threshold is conservative in the face
of possible incompleteness.

Figure~\ref{con} shows the resulting contour plot excluding various
dark-matter models at various confidence ($\sigma$) levels.
Finally, in Figure~\ref{comb}
we compare our 95\% confidence ($2\sigma$) limits
with those from the EROS \citep{af} and MACHO \citep{al4} 
microlensing collaborations.
The halo binaries exclude models that have generally higher MACHO
masses than those probed by these microlensing experiments.
Moreover, they extend all the way to the \citet{lo} limit based on stability
of the Galactic disk.

All three curves in Figure~\ref{comb} show limits on models with 
$\delta$-function mass distributions. However any model with broader 
distribution of mass but total density $\rho_H$, would be ruled out
provided that the combined limits from the three curves ruled out each mass
separately. 

Hence, the only remaining window open for MACHOs would be
black holes with a mass function strongly peaked at $M\sim35~\msun$.
Even this model is more strongly constrained than shown Figure~\ref{comb}
because the MACHO \citep{al4} results and our results each separately
place weak limits on this model, which when combined comes close to
ruling it out.

\section{Conclusions}
\label{co}
In this paper, we have investigated the evolution of halo wide binaries
in the presence of MACHOs and estimated upper limits of MACHO density as
a function of their assumed mass by comparing our simulations to the
sample of halo wide binaries of CG. We exclude MACHOs with masses $M>43\msun$
at the standard local halo density $\rho_H$ at the 95\% confidence level.

MACHOs have been a major dark-matter candidate ever since observations
first established that this mysterious substance dominates the mass of
galaxies. Prodigious efforts over several decades have gradually
whittled down the mass range allowed to this dark-matter candidate.
However, the window for MACHOs with $30\msun\lesssim M\lesssim10^3\msun$
remained completely open while constraints in the range 
$10^3\msun\lesssim M\lesssim10^6\msun$ were somewhat model dependent.
Our new limits on MACHOs $M>43\msun$ all but close this window.

\acknowledgments
We are grateful to \'Eric Aubourg and Kim Griest for providing data
for Figure~\ref{comb}. We thank John Bahcall, Bohdan Paczy\'nski, and
especially Scott Tremaine for valuable comments that significantly improved
the paper. A detailed critique by referee Terry Oswalt also greatly
improved the paper. This work was supported by grant AST 02-01266 from the NSF.

{
\appendix
\section{Ionized Binaries}
\label{app:ion}
After disruption of a binary system, the two ionized members remain in similar
Galactic orbit, so it is only the separation along the direction of the orbital
motion that can keep increasing, while the perpendicular separation oscillates.

In the Coulomb regime, the average post-ionization gain in the relative
velocity of binaries along the 
orbital direction is (see eq.[\ref{eq:cou}]),
\begin{equation}
\sqrt{\left<v^2_\parallel\right>}=
\sqrt{{32\pi G^2\rho M\Delta t\over 3v}\ln\Lambda},
\label{parallel}
\end{equation}
where $\Delta t$ is the remaining time to 10~Gyr after disruption. For
a diffusive process that is uniform over time $\Delta t$, the 
root-mean-square separation parallel to the orbital motion is
\begin{equation}
\sqrt{\left<d_\parallel^2\right>}=\sqrt{\left<v^2_\parallel\right>{\Delta t^2
\over3}}=2000~{\rm pc}\left({\rho\over\rho_H}\right)^{1/2}
\left(M\over 30M_\odot\right)^{1/2}
\left({v\over 300~{\rm km\ s^{-1}}}\right)^{-1/2}
\left({\ln\Lambda\over5.1}\right)^{1/2}
\left({\Delta t\over10~{\rm Gyr}}\right)^{3/2},
\label{pdist}
\end{equation}
where the Coulomb logarithm is calculated at the scaled quantities.

For a conservative limit of the tidal radius, $a_t=3$~pc,
the time required for ionized binaries to separate farther than $a_t$ 
is 0.13~Gyr, so that only binaries ionizing within the last $\sim$ 1\%
of the age of the Galaxy have a significant chance to be confused with
bound systems, even for the lowest mass perturbers that we can effectively
probe. In the tidal limit, binaries escape with characteristic velocities of
the transition separation $\sim10^4$~AU, i.e., 300~$\rm m\ s^{-1}$.
Hence, they drift one tidal radius in only 10~Myr, so their impact is
even smaller. Nevertheless, since there are only a handful of binaries
in the widest-separation bins, it is important to make a careful estimate
of the contribution from ionizing binaries. We take account of ionized binaries
in the simulations as follows. Binaries are considered ionized either when
they have positive energy or they have $a>a_t$. They are then assigned a
relative velocity equal to their escape velocity in the former case, or zero 
in the latter. A random orbital direction is chosen. The ionized binaries
in the Coulomb regime then continue to suffer perturbations to the end 
of the simulations whose effect we calculate using 
equation~(\ref{parallel}) and (\ref{pdist}).
At the end of the simulation, the binary is assigned a transverse separation
drawn randomly from a sinusoidal distribution of amplitude,
\begin{equation}
d_{\perp,{\rm MAX}}={v_\perp\over\Omega}=26~{\rm pc}~
\left({v_\perp\over{\rm km\ s^{-1}}}\right),
\end{equation}
where $v_\perp$ is the final transverse velocity, $\Omega=\sqrt{2}v_c/R_0$
is the epicyclic frequency, and $R_0$ is the Galactocentric distance.
Finally, we ``observe'' the ionized binary from a random orientation and
record the projected separation. We find that the ionized binaries have no
significant effect either on the final distribution or the calculated
likelihoods. We ignore ionized binaries in the tidal regime because these
escape with much higher initial velocities and because these velocities grow
much more rapidly than indicated by equation~(\ref{parallel}).
}


\begin{thebibliography}{99}
\frenchspacing

\bibitem[Afonso et al.(2003)]{af}
Afonso, C., et al. 2003, A\&A, 400, 951

\bibitem[Alcock et al.(1998)]{al0}
Alcock. C., et al. 1998, ApJ, 499, L9

\bibitem[Alcock et al.(2001)]{al4}
Alcock. C., et al. 2001, ApJ, 550, L169

\bibitem[Bahcall, Hut \& Tremaine(1985)]{bht}
Bahcall, J. N., Hut, P., \& Tremaine, S. 1985, ApJ, 290, 15

\bibitem[Bahcall \& Soneira(1980)]{lh}
Bahcall, J. N., \& Soneira, R. M. 1980, ApJS, 44, 73

\bibitem[Binney \& Tremaine(1987)]{gd}
Binney, J., \& Tremaine, S. 1987, Galactic Dynamics (Princeton: Princeton 
Univ. Press)

\bibitem[Chanam\'e \& Gould(2003)]{cg}
Chanam\'e, J., \& Gould, A. 2003, ApJ, in press

\bibitem[Chiba \& Beers(2000)]{cb}
Chiba, M., \& Beers, T. C. 2000, AJ, 119, 2843

\bibitem[De R\'ujula, Jetzer \& Mass\'o(1992)]{der}
De R\'ujula, A., Jetzer, P., \& Mass\'o, E. 1992, A\&A, 254, 99

\bibitem[Gould \& Salim(2003)]{sa1} 
Gould, A., \& Salim, S. 2003, ApJ, 582, 1001

\bibitem[Heggie(1975)]{heg}
Heggie, D., C. 1975, MNRAS, 173, 729

\bibitem[Lacey \& Ostriker(1985)]{lo}
Lacey, G., C., \& Ostriker, J. P. 1985, ApJ, 299, 633

\bibitem[Luyten(1979)]{nltt1}
Luyten, W. J. 1979-1980. New Luyten Catalogue of Stars with Proper
Motions Larger than Two Tenths of an Arcsecond (Minneapolis: Univ. Minnesota
Press)

\bibitem[Luyten \& Hughes(1980)]{nltt2}
Luyten, W. J., \& Hughes, H. S. 1980, Proper Motion Survey with the 
Forty-Eight Inch Schmidt Telescope, LV, First Supplement to the NLTT Catalogue
(Minneapolis: Univ. Minnesota Press)

\bibitem[Moore(1993)]{mo}
Moore, B. 1993, ApJ, 413, L93

\bibitem[Nemiroff et al.(1993)]{ne1}
Nemiroff, R. J., et al. 1993, ApJ, 414, 36

\bibitem[Odenkirchen et al.(2003)]{pal5}
Odenkirchen M., et al. 2003, AJ, in press

\bibitem[Popowski \& Gould(1998a)]{pg1}
Popowski, P., \& Gould, A. 1998a, ApJ, 506, 259

\bibitem[Popowski \& Gould(1998b)]{pg2}
Popowski, P., \& Gould, A. 1998b, ApJ, 506, 271

\bibitem[Retterer \& King(1982)]{rk}
Retterer, J. M., \& King, I. R. 1982, ApJ, 254, 214

\bibitem[Salim \& Gould(2003)]{sa2}
Salim, S., \& Gould, A. 2003, ApJ, 582, 1011

\bibitem[Weinberg, Shapiro \& Wasserman(1987)]{martin}
Weinberg, M. D., Shapiro, S. L., \& Wasserman, I. 1987, ApJ, 312, 367

\bibitem[Zheng et al.(2001)]{zh}
Zheng, Z., Flynn, C., Gould, A., Bahcall, J. N., \& Salim, S. 2001, ApJ, 555,
393

\end{thebibliography}
\end{document}